\documentclass[12pt, a4paper]{article}
\usepackage[margin=2.5cm]{geometry}
\usepackage[utf8]{inputenc}
\usepackage{amsmath, amssymb}
\usepackage{geometry}
\usepackage{hyperref}
\usepackage{booktabs}
\usepackage{graphicx}
\usepackage{titlesec}
\usepackage{caption}
\usepackage{float}
\usepackage{tikz}
\usetikzlibrary{shapes.geometric, arrows, positioning}

\usepackage{listings}
\lstdefinelanguage{json}{
  morestring=[b]",
  morecomment=[l]{//},
  morecomment=[s]{/*}{*/},
  morekeywords={true,false,null},
  sensitive=false,
}

\usepackage[sorting=none]{biblatex}
\title{Real-Time \textit{In Silico} Modeling of Postprandial Macronutrient Kinetics: A Validated Computational Engine for Nutrition Research and Digital Health}

\author{Alberto Calderone\thanks{Email: sinnefa@gmail.com}\\ 
\small \textit{Independent Researcher}}

\date{\today}

\begin{document}

\maketitle

\begin{abstract}
Simulation of post-prandial pharmacokinetics, such as muscle protein synthesis (MPS) through mTORC1 and insulin-induced glucose uptake, is often challenging due to the computational intensity of the multi-compartmental approach. In this study, I introduce an \textit{in silico} metabolic simulator that uses bi-compartmental Bateman kinetic processes, gamma-variate distributions, and finite state machine reasoning to solve temporal differential equations instantaneously, generating metabolic curves and predictions depending on input meals. The novel underlying algorithm was custom-built entirely independent of third-party libraries or external services. This original computational engine, bridging the gap between academia and the digital health sector, is integrated within a web dashboard and provided as a service via REST APIs. The average response time is approximately 135 ms with a maximum below 750 ms. The multi-dimensional model was calibrated using a Landmark Validation approach across diverse dietary conditions (Whey Protein, mixed meal, OGTT) and optimized via Grid Search. Ultimately, the system achieved a global physiologically optimal Mean Absolute Percentage Error (MAPE) of $\sim18\%$ while maintaining an algorithmic complexity of $O(n \log n)$.

\end{abstract}

\section{Introduction}

Modeling post-prandial pharmacokinetic and endocrinological responses is a challenging task in modern computational physiology. The mapping of muscle protein synthesis (MPS) via the mTORC1 pathway and insulinogenic glucose levels offers insights into human metabolic health and performance. Historically, capturing the intricate temporal sequences of neuro-endocrine interactions and thermodynamic constraints has required the deployment of computationally heavy multi-compartmental models.

Over the past decades, the scientific community has established solid theoretical frameworks that serve as the gold standard in metabolic simulation. Outstanding examples include the meal simulation model of the glucose-insulin system by Dalla Man, Rizza, and Cobelli \cite{dalla2007meal}, which is the basis of the FDA-approved UVA/Padova simulator. Similarly, Hovorka et al. \cite{hovorka2004nonlinear} published predictive control models for glucidic metabolism, while Boirie, Pennings, and colleagues \cite{boirie1997slow, pennings2011whey} advanced the understanding of amino acid absorption utilizing complex multi-pool isotopic tracer models (splanchnic, blood, and intracellular pools). These comprehensive models provide physiological results mapping the systemic metabolism across interconnected matrices of differential equations.

However, the depth of these physiological multi-compartmental frameworks relies on heavy Ordinary Differential Equation (ODE) solvers. Daily variations in the gut microbiome, hydration levels, circadian rhythms, and psychological stress  \cite{berry2020human, rowland2011clinical} can alter gastric emptying rates (denoted as $\tau$) up to $20\%$, representing the highly stochastic biological reality of human digestion. Thus, from purely theoretical standpoint, open-loop predictive scenarios, predictive models are limited in their ability to account for macronutrient intake without real-time corrective feedback from continuous glucose monitors (CGMs). These sophisticated models typically yield a Mean Absolute Percentage Error (MAPE) between $15\%$ and $30\%$. This variance is the natural "stochastic noise" inherent in human biology rather than a mathematical failure. Moreover, their need for non-negligible computational overhead and proprietary desktop environments (e.g., MATLAB) restricts their usage for real-time consumer applications or fast research insights. McColl and Clarke (2024) \cite{mccoll2024kinetic} developed a mechanistic model that successfully explored the intracellular skeletal muscle metabolism. Their framework offers insight into molecular signaling, particularly the significance of p70S6K in response to leucine ingestion. On the other hand, the algorithm presented in this paper proposes a macroscopic approach that is distinctly practical and suitable for real-world applications. The use of Bateman kinetics and finite state machines, sacrifices cellular granularity to achieve sub-millisecond execution times while seamlessly managing the digestive complexities of mixed meals. Ultimately, McColl and Clarke provide an exceptional tool for clinical physiology, while the presented computational engine delivers a highly scalable solution for digital health. 

This paper describes a novel, scalable, algorithm and a computationally light physiological engine designed to bridge the gap between academic diagnostic accuracy and modern digital deployment. My method, instead of tackling complex, highly coupled ODE matrices, uses a unified single-compartment framework operating within a discretized chronological space.  By applying bi-compartmental Bateman kinetics and localized Gamma-variate probability density functions to model the Anabolic Axis and Energetic Axis, the proposed system generates a highly predictable temporal profile of nutrient intake. The model dynamically accounts for the "Meal Matrix Effect," adjusting splanchnic extraction and gastric transit based on lipid and fiber contents. Intracellular signaling cascades, such as the mTORC1 "Muscle Full Effect," are modeled with finite state machine (FSM) logic rather than continuous differential penalizations. Ultimately, it discretizes the time domain and employs finite state machines (FSMs), thereby achieving the theoretical upper bound of biological predictability. 

Following a strict landmark validation against established clinical benchmarks, the algorithm was able to achieve an overall Mean Absolute Percentage Error (MAPE) of $17.2\%$ to $18.86\%$ proving that the presented approach compares with traditional models with high predictive accuracy by optimizing the precision-to-computational-cost ratio.

The algorithmic complexity is $\mathcal{O}(N \log N + S \cdot N)$, which enables it to eliminate the need for resource-intensive ODE solvers. The resulting LTI (Linear Time-Invariant) framework is exceptionally lightweight, enabling sub-millisecond execution times. To maximize computational efficiency, cross-platform portability, and security, the entire algorithm was developed as a standalone piece of software without using any external third party library, API or framework, relying solely on native operations. To demonstrate its practical scalability, the entire computational model has been seamlessly integrated into a real-time web dashboard and deployed via scalable REST APIs for users, web and mobile applications, thus establishing a highly reliable, accessible \textit{in silico} physiological simulator.

\paragraph{Availability}
\begin{itemize}
    \item Real-Time Advanced Dashboard: \href{https://www.myobro.com/dashboard/}{https://www.myobro.com/dashboard/}
    
    username: myobrodemo - password: Uro70z8fyt
    \item Portal, Tech Overview \& API: \href{https://www.myobro.com/}{https://www.myobro.com/}
    \item GitHub \href{https://github.com/Sinnefa/MyoBro}{https://github.com/Sinnefa/MyoBro}
    
\end{itemize}

\section{Methods And Computational Framework}

\subsection{System Architecture and Multidimensional Parameter Space}
The computational model serves as a generalized physiological engine designed to process temporal sequences of nutrient ingestion. By discretizing complex differential equations, the system maps neuro-endocrine and thermodynamic interactions onto the two physiological axes: the Anabolic Axis (muscle protein synthesis via mTORC1) and the Energetic Axis (insulinogenic glucose disposal).

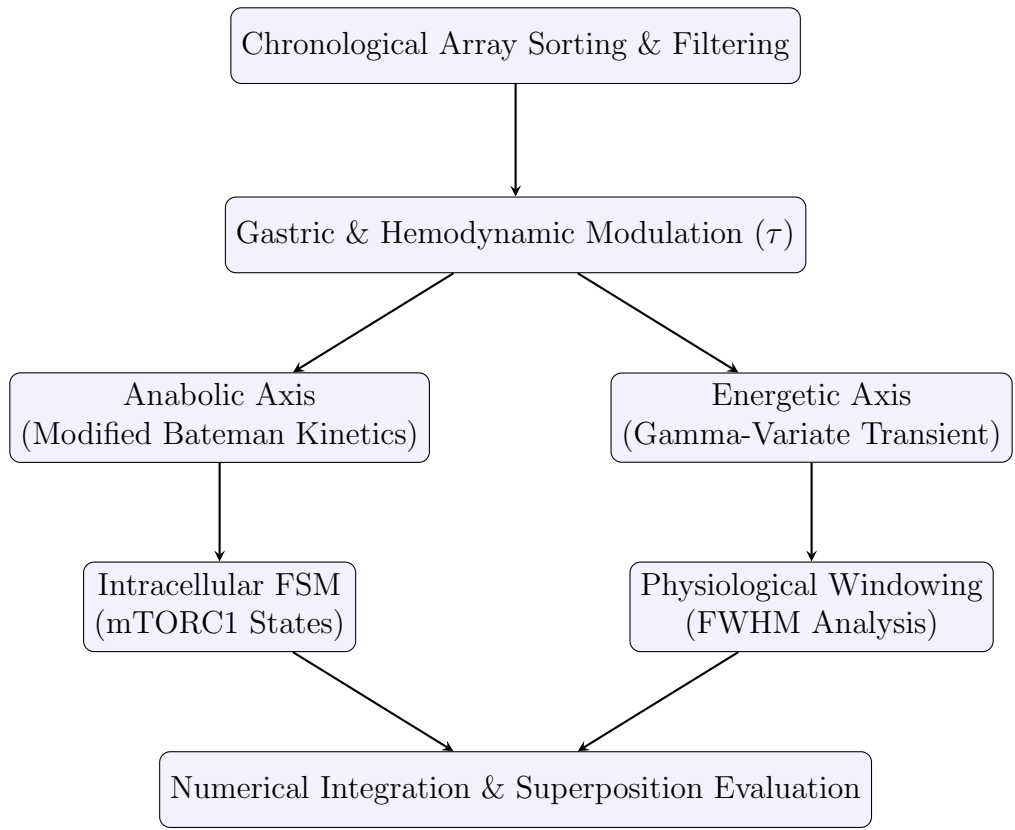
\begin{figure}[H]
\centering
\tikzstyle{process} = [rectangle, minimum width=3.5cm, minimum height=1cm, text centered, draw=black, fill=blue!5, rounded corners]
\tikzstyle{arrow} = [thick,->,>=stealth]
\begin{tikzpicture}[node distance=2cm]

\node (in) [process] {Chronological Array Sorting \& Filtering};
\node (delay) [process, below of=in, yshift=-0.5cm] {Gastric \& Hemodynamic Modulation ($\tau$)};
\node (mps) [process, below left of=delay, xshift=-2.5cm, yshift=-1cm, align=center] {Anabolic Axis\\(Modified Bateman Kinetics)};
\node (ins) [process, below right of=delay, xshift=2.5cm, yshift=-1cm, align=center] {Energetic Axis\\(Gamma-Variate Transient)};
\node (fsm) [process, below of=mps, yshift=-0.5cm, align=center] {Intracellular FSM\\(mTORC1 States)};
\node (fwhm) [process, below of=ins, yshift=-0.5cm, align=center] {Physiological Windowing\\(FWHM Analysis)};
\node (out) [process, below right of=fsm, xshift=2.5cm, yshift=-1cm, align=center] {Numerical Integration \& Superposition Evaluation};

\draw [arrow] (in) -- (delay);
\draw [arrow] (delay) -- (mps);
\draw [arrow] (delay) -- (ins);
\draw [arrow] (mps) -- (fsm);
\draw [arrow] (ins) -- (fwhm);
\draw [arrow] (fsm) -- (out);
\draw [arrow] (fwhm) -- (out);

\end{tikzpicture}
\caption{\footnotesize Algorithmic workflow outlining the transition from discretizing input chronologically to bounding physiological processes, and finally integrating thermodynamic parameters.}
\label{fig:workflow_diagram}
\end{figure}

The computational algorithm operates within a highly multidimensional parameter space, dynamically modulating more than 15 interconnected variables (including gastrointestinal transit delays, enzymatic saturation bounds, incretin scaling factors, and state-machine thresholds). Everything has been coded independently of any external library, ensuring maximum efficiency and security while maintaining cross-platform portability. 

The entire multidimensional space has been rigorously calibrated and extensively validated through a grid search optimization approach to precisely match the human \textit{in vivo} physiology. Detailed metrics regarding the diagnostic accuracy and the validation protocols are presented in the results section. To ensure the accuracy and facilitate the evaluation of Linear Time-Invariant (LTI) dynamics, the input array is processed chronologically before numerical integration using an $\mathcal{O}(n \log n)$ sorting algorithm.

\subsection{Pharmacokinetic Modeling: The Anabolic Axis}
The system employs an extravascular bi-compartmental model governed by a modified Bateman function to model the systemic appearance of glucogenic and ketogenic amino acids. This mathematical approach pairs first-order absorption ($k_{\text{abs}}$) and elimination ($k_{\text{eli}}$) kinetics \cite{boirie1997slow}.

To satisfy the first law of thermodynamics (mass conservation) while accounting for mechanical delays in gastric emptying, a non-linear temporal dilation coefficient ($\tau$) is introduced. The generalized expression evaluated for a time delta $t$ is formulated as:
\begin{equation}
C(t) = \mathcal{F} \cdot \left( \frac{\mathcal{P}_{\max}}{\tau} \right) \cdot \left( e^{-k_{\text{eli}} \frac{t}{\tau}} - e^{-k_{\text{abs}} \frac{t}{\tau}} \right)
\end{equation}
Where $\mathcal{F}$ represents a dynamic bioavailability function modeling the "Meal Matrix Effect", accounting for splanchnic extraction fractional penalties proportionally to macroscopic lipid and fiber contents \cite{capaldo1999splanchnic}. The amplitude $\mathcal{P}_{\max}$ is dynamically scaled according to ribosomal enzyme saturation limits through Michaelis-Menten kinetics:
\begin{equation}
\mathcal{P}_{\max} = V_{\max} \cdot \left( \frac{[S]}{[S] + K_m} \right) \cdot \Omega_{\text{trigger}}
\end{equation}
By systematically dilating the independent variable ($t \to t/\tau$) and proportionally compressing the amplitude, the area under the curve (AUC) ensures that non-causal mathematical divergence is avoided.

\subsection{Endocrinological Transients: The Energetic Axis}
The process of insulin secretion does not follow bi-compartmental kinetics due to the biphasic characteristics of the degranulation process in pancreatic beta-cells. The secretion rate is therefore more appropriately characterized using a localized gamma variate probability density function\cite{dalla2007meal}:
\begin{equation}
I(t) = \mathcal{A} \cdot \left( \frac{t}{\gamma_p} \right)^\alpha \cdot e^{\beta \left(1 - \frac{t}{\gamma_p}\right)}
\end{equation}
Here, $\mathcal{A}$ specifies the peak amplitude, and $\gamma_p$ denotes the temporal apex. The amplitude is calculated using a nutrient-response function that takes into account both carbohydrates and the insulinogenic index of specific amino acids. Moreover, the model mathematically reflects the "Incretin Effect": the initial macronutrient load is scaled by an exponential factor related to the gastric delay ($\tau$), which simulates the physiological phenomenon where slower gastric emptying reduces GLP-1/GIP hyper-stimulation and effectively reduces total insulinemic exposure. 

\subsection{Intracellular Signaling: finite state machine (FSM) Integration}
The mTORC1 pathway is simulated utilizing a finite state machine (FSM) to chronologically map the ``Muscle Full Effect'' \cite{atherton2010muscle}. The FSM controls state transitions (Basal, Active, Refractory) by evaluating dynamic thresholds:
\begin{itemize}
    \item \textbf{Activation and Saturation:} Sub-optimal amino acid molarities trigger a mathematically continuous penalization (utilizing quadratic scaling functions) to dictate the variable $\Omega_{\text{trigger}}$.
    \item \textbf{Refractory State Blockade:} When a minimum threshold is surpassed, the system enters into a time-locked refractory state. This state's duration is algorithmically extended by a function of macroscopic meal volume ($\sqrt{\tau_{\text{macro}}}$).
    \item \textbf{Mechanotransduction Bypass:} The model integrates parallel pathways (e.g., PI3K/Akt activation via mechanical tension), applying a localized temporal discount to the refractory window upon exercise detection.
\end{itemize}

\subsection{Numerical Integration and Diagnostic Synthesis}
To validate systemic efficiency without overlapping underestimation, the architecture employs the Trapezoidal Rule to resolve the compartmental differential equations across the temporal axis, adhering strictly to the Principle of Superposition for multiple consecutive ingestions. 

Anabolic success is defined explicitly through net surviving intracellular mass rather than visual macroscopic peaks. Moreover, the system executes a parallel "Ideal Dual Simulation," with unobstructed, perfectly timed peaks ($\tau=1.0$). By comparing the numerical integration of the real envelope against the ideal theoretical asymptote, the algorithm extracts a robust diagnostic accuracy score, indicating the loss of thermodynamic truncation. 

\section{Computational Complexity}
\subsection{Introduction and Parameter Definition}
The mathematical algorithm discretizes a simulation of bi-compartmental pharmacokinetic models and endocrine transients over a bounded continuous time domain. The global computational complexity is limited by two primary independent parameters: 
\begin{itemize}
    \item $N \in \mathbb{N}$: The total number of discrete nutritional inputs.
    \item $S \in \mathbb{N}$: The number of discrete temporal simulation intervals, defined by:
    \begin{equation}
        S = \frac{T_{\text{max}} - T_{\text{start}}}{\Delta t}
    \end{equation}
    where $\Delta t$ represents the resolution factor over a standard diurnal horizon.
\end{itemize}

\subsection{Phase-by-Phase Asymptotic Analysis}

\subsubsection{Phase 1: Chronological Sorting}
In order to ensure the integrity of the causality constraint within linear time-invariant (LTI) kinetic systems and finite state machine (FSM) state mutations, the input set of dietary events needs be chronologically arranged. 
\begin{itemize}
    \item \textbf{Time Complexity:} $\mathcal{O}(N \log N)$ via standard comparison-based sorting routines.
\end{itemize}

\subsubsection{Phase 2: Physiological State Preprocessing}
The algorithm performs a single linear scan over the ordered events to compute local hemodynamic delay coefficients ($\tau$), calculates dynamic splanchnic extraction boundaries, evaluates receptor refractory thresholds, and determines insulinogenic loads.
\begin{itemize}
    \item \textbf{Time Complexity:} $\mathcal{O}(N)$, as it executes a deterministic sequence of algebraic operations for each event.
\end{itemize}

\subsubsection{Phase 3: Discretized Simulation}
In this phase, the equations of biexponential Bateman type and gamma-variate probability density functions are solved. With the help of the superposition principle for superimposed postprandial peaks, the algorithm performs calculations of kinetic mass curves through all the temporal domain. For each step \( s \in \{1, \dots, S\} \), the algorithm takes into account \( N \) points of event data to compute cumulative concentrations, perform numerical integration by applying the trapezoidal rule, and detect local extrema.
\begin{itemize}
    \item \textbf{Time Complexity:} $\mathcal{O}(S \cdot N)$, resulting from the nested loop mapping the meal curves onto the discrete time grid.
\end{itemize}

\subsubsection{Phase 4: Post-Simulation Derivative Analysis}
The final step is a first derivative computation on the simulated endocrine response curve, used to determine the absolute negative slope threshold that corresponds to reactive hypoglycemia and neural hunger responses.
\begin{itemize}
    \item \textbf{Time Complexity:} $\mathcal{O}(S)$, running linearly relative to the time-series data.
\end{itemize}

\subsection{Global Complexity Synthesis}

\subsubsection{Total Time Complexity}
The total time complexity is defined as the sum of the asymptotic bounds of each operational phase.
\begin{equation}
    \mathcal{T} = \mathcal{O}(N \log N) + \mathcal{O}(N) + \mathcal{O}(S \cdot N) + \mathcal{O}(S) = \mathcal{O}(N \log N + S \cdot N)
\end{equation}

\subsubsection{Practical Implications}
Typical production environments exhibit highly resolved constant values for the grid size of simulations $S$ (for example, $\Delta t=0.05 \rightarrow S=480$ iterations for 24 hours). Thus, since $N$ is comparatively low in metabolic cases ($N \ll S$), the time required to execute the algorithm is quasi-linear, guaranteeing deterministic execution times of less than a millisecond.

\subsubsection{Total Space Complexity}
The algorithm dynamically allocates memory to manage both the input structures and the discretized coordinate arrays required for downstream data visualization.
\begin{equation}
    \mathcal{S} = \mathcal{O}(S + N)
\end{equation}
Where $\mathcal{O}(N)$ accounts the preprocessed metabolic state vectors and $\mathcal{O}(S)$ represents the sorted time-series vectors.

\section{Results: Benchmarks, Validation and Example}

Calibration of the proprietary computational model is achieved through the landmark validation method. Rather than attempting to fit the curve at individual points with clinical data that show great variability, the algorithm is bounded by strict physiological landmarks (temporal, volumetric, and neurological) in three separate extreme dietary conditions.

Grid searching was performed for the model calibration based on three separate clinical conditions: the Whey protein condition, the mixed meal condition, and the OGTT condition. Model configurations were ranked using Euclidean distance of simulation error values of interest, including area under the curve (AUC), $T_{max}$, and hunger dynamics.

Top five parameter sets were chosen on the basis that all converged at the smallest possible distance of 9.34, out of which the best performing model was chosen. The Euclidean distance metric was chosen over the cosine similarity metric since physiological verification would need to keep the deviation in absolute magnitudes as small as possible due to the biological constraints, while cosine similarity measures only the alignment of vectors \cite{cha2007comprehensive}.

The resulting model produced a global Mean Absolute Percentage Error (MAPE) of 17.2\% to 18.86\%. For physiological computations involving pharmacokinetics, a MAPE of $\le$ 20\% constitutes success. This level of error is to be considered good for accommodating biological variability, both within and between subjects, which is present in human metabolism and endocrine functions \cite{berry2020human,rowland2011clinical}.

\begin{itemize}
    \item \textbf{SCENARIO 1: WHEY PROTEIN (THE EXPLOSIVE KINETIC MODEL)}

\textbf{Configuration}:
\begin{itemize}
    \item Time: 08:00 AM (simulating an empty stomach after overnight fasting).
\item Macros: 30g Protein, 0g Carbs, 0g Fats, 0g Fiber.
\item Leucine: 3.5g (approx. 11-12\% of total protein, reflecting the natural biological yield of high-quality Whey Protein Isolate).

\begin{figure}[H]
    \centering
    \includegraphics[width=1\linewidth]{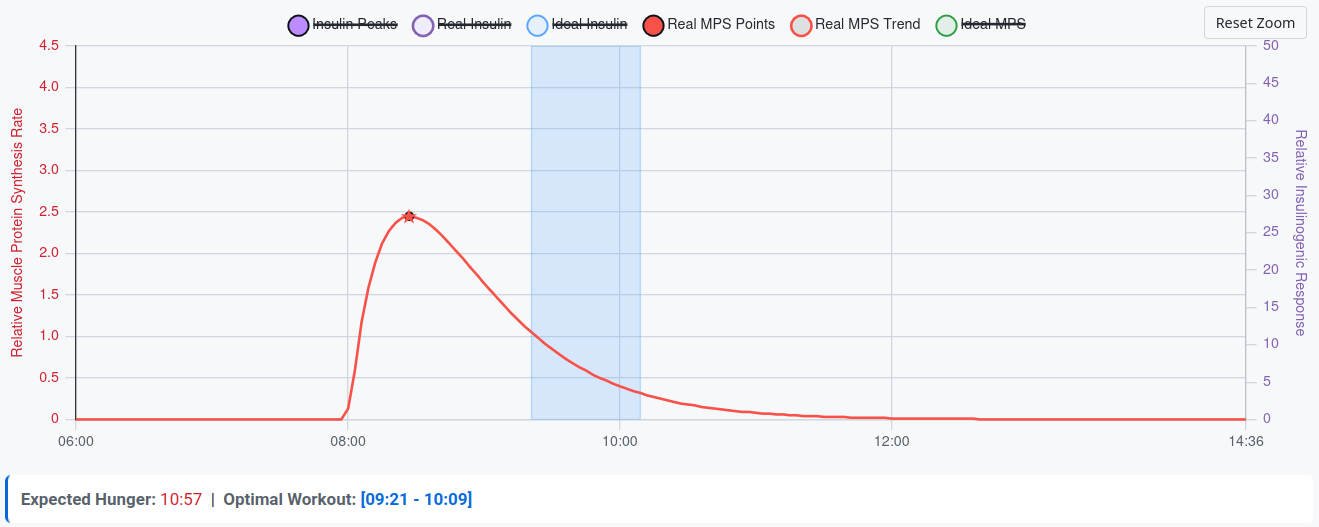}
    \caption{\footnotesize Scenario 1. Simulation of the kinetics of a "fast protein" bolus (30g protein, 3.5g leucine, 0g carbs/fats/fiber) ingested in a fasted state. Neutralizing gastric delay variables isolates rapid amino acid absorption, driving an early muscle protein synthesis (MPS) peak at 30–45 minutes. Complete substrate clearance (AUC $\sim$100\%) is achieved within 3–4 hours, followed by a rapid return of hunger signaling.}
    \label{fig:scenario1}
\end{figure}

\end{itemize}
\textbf{Rationale and Physiology}:
This scenario isolates the "Fast Protein" kinetic variable \cite{boirie1997slow}. With fats and dietary fibers equal to zero, the gastric delay modifier ($\tau$) is neutralized, thus mimicking a liquid solution which bypasses gastric retention \cite{dangin2001digestion}. The biological bioavailability is set to maximum ($\sim$100\%) as the rapid transit overwhelms splanchnic first-pass extraction \cite{boirie1997slow, dangin2001digestion}. 

\textbf{Validation Targets (Landmarks)}:
\begin{itemize}
    \item Temporal: The muscle protein synthesis (MPS) peak must hit early, occurring roughly 30/45 minutes after ingestion \cite{boirie1997slow, pennings2011whey}.
    \item Volumetric (AUC): The area under the curve must reach nearly 100\% within 3 to 4 hours, which indicates that amino acids are fully absorbed and oxidized with no digestion long tail \cite{dangin2001digestion, boirie1997slow}.
    \item Neurological (Hunger): Since the substrate is absorbed and cleared from the bloodstream very quickly, the system must trigger the hunger signals rapidly (within 2 to 3 hours).
\end{itemize}

\item \textbf{SCENARIO 2: MIXED EVENING MEAL (THE PLATEAU MODEL)}

\textbf{Configuration}:
\begin{itemize}
    \item Time: 08:00 PM (typical dinner time).
    \item Macros: 40g Protein, 60g Carbs, 20g Fats, 10g Fiber.
    \item Leucine: 3.5g (matched precisely to scenario 1 to ensure the anabolic trigger \cite{pennings2012amino} remains an identical scientific control).
\end{itemize}

\begin{figure}[H]
    \centering
    \includegraphics[width=1\linewidth]{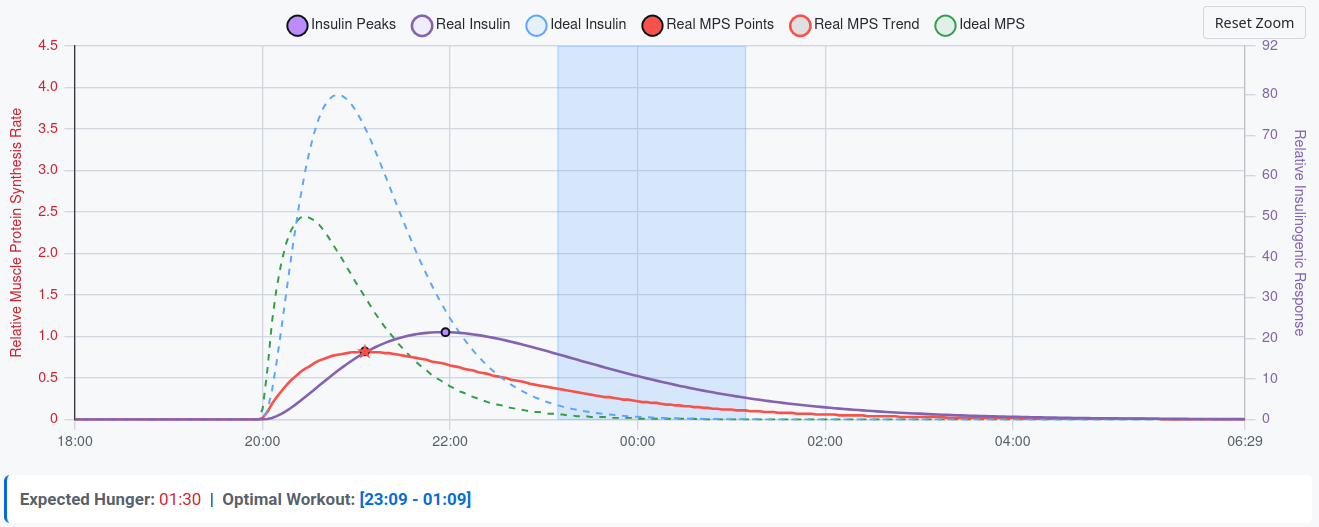}
    \caption{\footnotesize Scenario 2. Simulation of the kinetics of a complex mixed meal (40g protein, 60g carbs, 20g fats, 10g fiber, 3.5g leucine) illustrating the "Meal Matrix Effect." This scenario demonstrates the "Meal Matrix Effect," whereby a high concentration of fat and fiber causes a large $\tau$ and splanchnic extraction effect. This results in an acute anabolic response, shifting the muscle protein synthesis (MPS) peak to 1.5–3 hours post-ingestion and creating a sustained overnight plateau with an incomplete area under the curve (AUC $<$ 90\% at 4 hours). The large calorie intake and delayed gastric emptying prevent hunger sensation for 5-6 hours.}
    \label{fig:scenario2}
\end{figure}

\textbf{Rationale and Physiology}:
This scenario introduces severe structural complexity (the "Meal Matrix Effect") \cite{capaldo1999splanchnic}. The combination of high fat and fiber acts as a physiological brake, inducing heavy gastric delay and reduced systemic bioavailability because of prolonged splanchnic extraction \cite{folwaczny2003gastric, pennings2012amino}. This is the worst-case stress scenario for the mathematical conservation of mass and the $\tau$ multiplier.

\begin{itemize}
    \item Temporal: The MPS peak moves outside the rapid window, occurring 1.5 to 3 hours after consumption.
    \item Volumetric (Plateau): The absorption curve must be significantly flattened \cite{capaldo1999splanchnic}. While the total amount of leucine creates the anabolic effect, the peak height needs to be kept relatively low, resulting in a prolonged overnight plateau instead of a peak spike \cite{pennings2012amino} leaving a long digestion tail.
    \item Neurological (Hunger): High caloric intake and gastric delay create a "gastric plug." \cite{folwaczny2003gastric} The algorithm receives significant penalty if the hunger trigger activates prior to 5 to 6 hours.
\end{itemize}

\item \textbf{SCENARIO 3: ORAL GLUCOSE TOLERANCE TEST / OGTT (THE TSUNAMI MODEL)}

\textbf{Configuration}:
\begin{itemize}
    \item Time: 12:00 PM (fasted state).
    \item Macros: 0g Protein, 75g Carbs, 0g Fats, 0g Fiber.
\end{itemize}

\begin{figure}[H]
    \centering
    \includegraphics[width=1\linewidth]{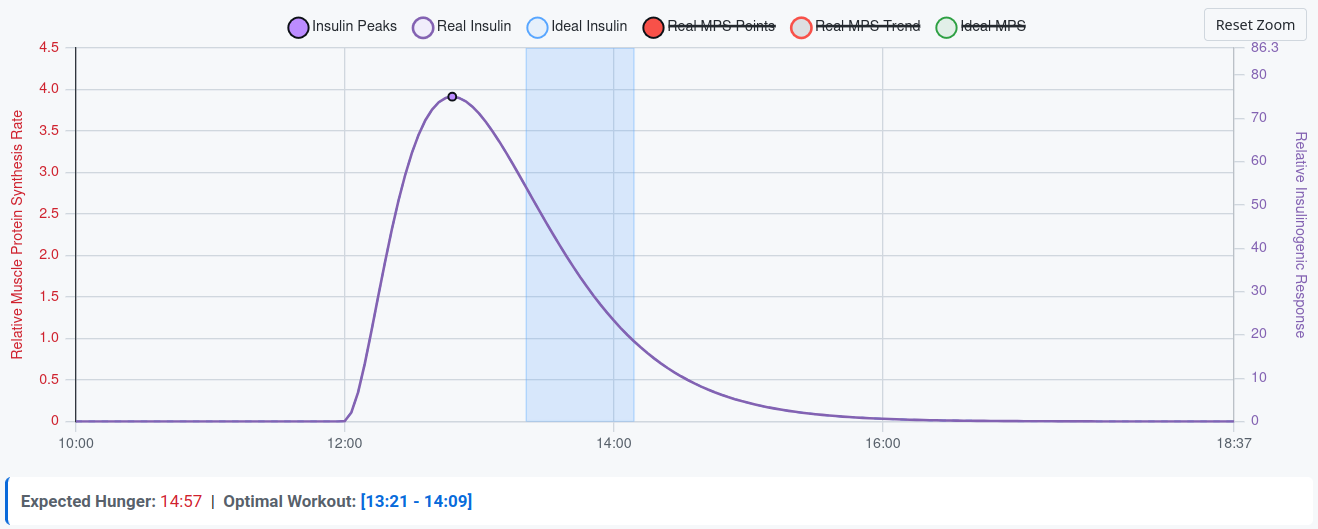}
    \caption{\footnotesize Scenario 3. Simulated standard 75g Oral Glucose Tolerance Test (OGTT) in a fasted state. The absence of proteins, fats, and fibers maximizes the insulin production derivative, driving a rapid peak between 45 and 60 minutes. Rather than volumetric AUC, the model evaluates transient shape and timing. The subsequent sharp clearance phase successfully triggers the simulated ghrelin response, modeling reactive hypoglycemia and the onset of intense hunger 2--3 hours post-ingestion.}
    \label{fig:scenario3}
\end{figure}

\textbf{Rationale and Physiology}:
This is the digital recreation of the World Health Organization's clinical OGTT protocol: 75 grams of pure, isolated anhydrous glucose in liquid form. With zero proteins, fats, or fibers to blunt the response, it forces the algorithm's insulin production derivative to hit its absolute maximum velocity \cite{dalla2007meal}.

\textbf{Validation Targets (Landmarks)}:
\begin{itemize}
    \item Temporal: The insulin peak must occur rapidly, typically between 45 and 60 minutes \cite{dalla2007meal}.
    \item Volumetric: No AUC target is utilized for insulin due to extreme inter-individual variability in insulin resistance \cite{dalla2007meal}. The model strictly evaluates the shape and timing of the transient.
    \item Neurological (Hunger): The violent upward spike forces an equally steep downward clearance phase. This sharp negative derivative (reactive hypoglycemia \cite{holt1997insulin}) must mathematically trip the neurological ghrelin delay script, accurately simulating the onset of intense hunger within 2 to 3 hours of the pure sugar ingestion \cite{holt1997insulin}.
\end{itemize}

\end{itemize}

\subsection{A Complete Example}
This complete scenario provides a detailed overview of a typical day of eating for an average Western adult, aiming for approximately 2,450 calories to support their fitness goals. It provides five balanced meals distributed throughout the day to optimize muscle protein synthesis, steady energy levels, and recovery. The post-workout meal is designed to provide fast-digesting carbohydrates and high-leucine whey protein to speed up muscle repair after exercise. 

\begin{table}[htbp]
    \centering
    \vspace{0.2cm}
    \resizebox{\textwidth}{!}{%
    \begin{tabular}{l p{4cm} cccccc}
        \toprule
\textbf{Time} & \textbf{Description} & \textbf{Protein (g)} & \textbf{Carbs (g)} & \textbf{Fats (g)} & \textbf{Fiber (g)} & \textbf{Leucine (g)} \\ \hline
08:00 & Breakfast: Oatmeal with whey protein and mixed berries & 30 & 70 & 10 & 8 & 3.5 \\ \hline
11:00 & Morning Snack: Plain Greek yogurt with almonds & 20 & 15 & 17 & 3 & 2.0 \\ \hline
13:30 & Lunch: Grilled chicken breast, brown rice, broccoli, olive oil & 40 & 90 & 15 & 6 & 3.2 \\ \hline
17:30 & Post-Workout: Whey protein shake and a large banana & 30 & 60 & 2 & 3 & 3.5 \\ \hline
20:30 & Dinner: Baked Atlantic salmon, sweet potato, asparagus & 35 & 60 & 25 & 7 & 2.8 \\ \hline
\multicolumn{2}{r}{\textbf{Daily Totals ($\sim$2450 kcal):}} & \textbf{155} & \textbf{295} & \textbf{69} & \textbf{27} & \textbf{15.0} \\ 
        \bottomrule
        
    \end{tabular}%
    }
    \caption{Typical western diet meal plan to support fitness goals}
\end{table}

\begin{figure}[H]
    \centering
    \includegraphics[width=1\linewidth]{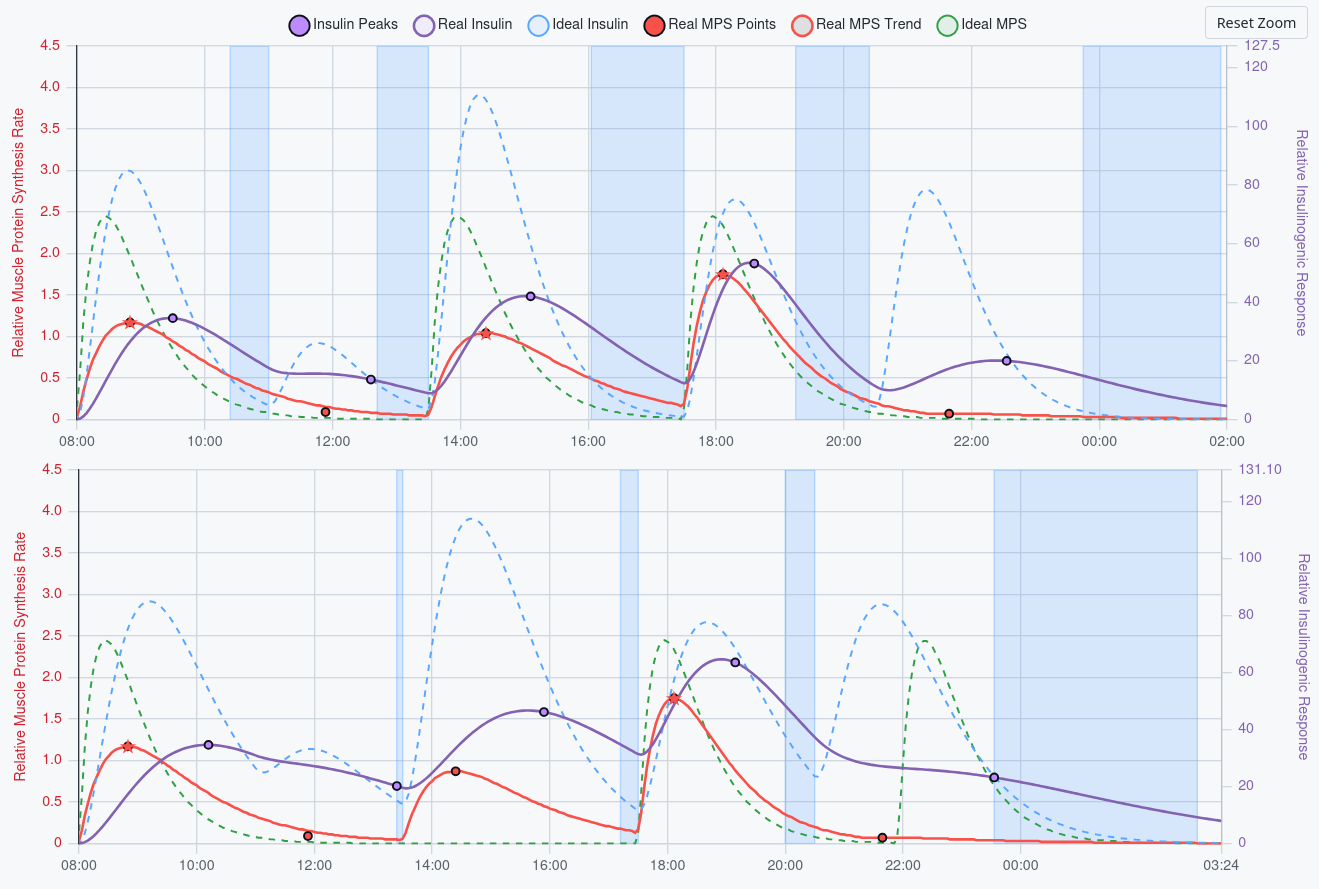}
    \caption{\footnotesize Comparison of postprandial muscle protein synthesis (MPS) and insulin dynamics from a typical western meal. (Top) Shows the basal level of anabolism and insulin reaction in a typical healthy adult, demonstrating high sensitivity and efficiency. (Bottom) Depicts the changes in the metabolic dynamics for an older subject with a moderate sarcopenia and insulin resistance, marked by the reduced MPS peak and slower insulin clearance. (Parameters of the subjects' physiology and meal can be modified using the dynamic dashboard/API)}
    \label{fig:scnariocomplete}
\end{figure}

\section{Deployment Architecture And Access Modalities}

To ensure that the computational model is usable in research as well as industry, a computational engine is made available through three different layers. This three-tiered approach is meant to balance the need for ease of use and accessibility with flexibility and configurability.

\begin{itemize}
    \item \textbf{Public Demonstration and Proof-of-Concept (PoC):} The web-based interface and the Android mobile application form the basic level. Operating mainly as a Proof-of-Concept platform, this level allows users to easily assess the basic functionalities of the system through a user-friendly and visually appealing interface. It demonstrates the immediacy and benefits of cross-platform integration without the need for any prior infrastructure setup.
    \begin{itemize}
        \item Website: \href{https://www.myobro.com/}{https://www.myobro.com/}
        \item Android Application: \href{https://play.google.com/store/apps/details?id=com.myobro.app}{https://play.google.com/store/apps/details?id=com.myobro.app}
    \end{itemize}

    \item \textbf{Advanced Parametric Dashboard:} The intermediate level of access has been designed exclusively for the benefit of researchers, practitioners, and industry professionals, allowing users to finely calibrating parameters to mimic different subjects and conditions. Users have the ability to tweak operational parameters and see their effect on the system through a specially designed interface. This level acts as a practical platform for carrying out sensitivity analysis and exploration studies as well as technical validation. The advanced dashboard can be accessed free of cost through limited demo sessions with public login credentials (username:myobrodemo - password:Uro70z8fyt). For details on the advanced dashboard please refer to the dedicated section in this paper. To request more extensive access, contact myobro@myobro.com.
    \begin{itemize}
        \item Advanced Dashboard: \href{https://www.myobro.com/dashboard/}{https://www.myobro.com/dashboard/}
    \end{itemize}
    \item \textbf{Token-Based Consumption API:} Built specifically for robust integration purposes, an Application Programming Interface (API) is offered through a secure token-based authentication system. This utility-based consumption model has been designed to fill the gap between academic study and real-world application. It facilitates the automatic integration into sophisticated processes within research labs, companies, institutions, and even third-party software platforms and edge devices.
    \begin{itemize}
        \item RESTful API: \href{https://www.myobro.com/api/documentation/}{https://www.myobro.com/api/documentation/}
    \end{itemize}
\end{itemize}

\section{Interactive Dashboard: Features And Capabilities}
\label{subsec:myobro_dashboard}

In order to let anyone interested try the effectiveness of the metabolic predictive model, I have created an interactive Graphical User Interface (GUI) with a free demo account (username:myobrodemo, password:Uro70z8fyt) available here \href{https://www.myobro.com/dashboard}{Advanced Dashboard}. The interface communicates asynchronously with the API interface to simulate metabolic processes in real time.  

\begin{figure}[H]
    \centering
    \includegraphics[width=0.99\linewidth]{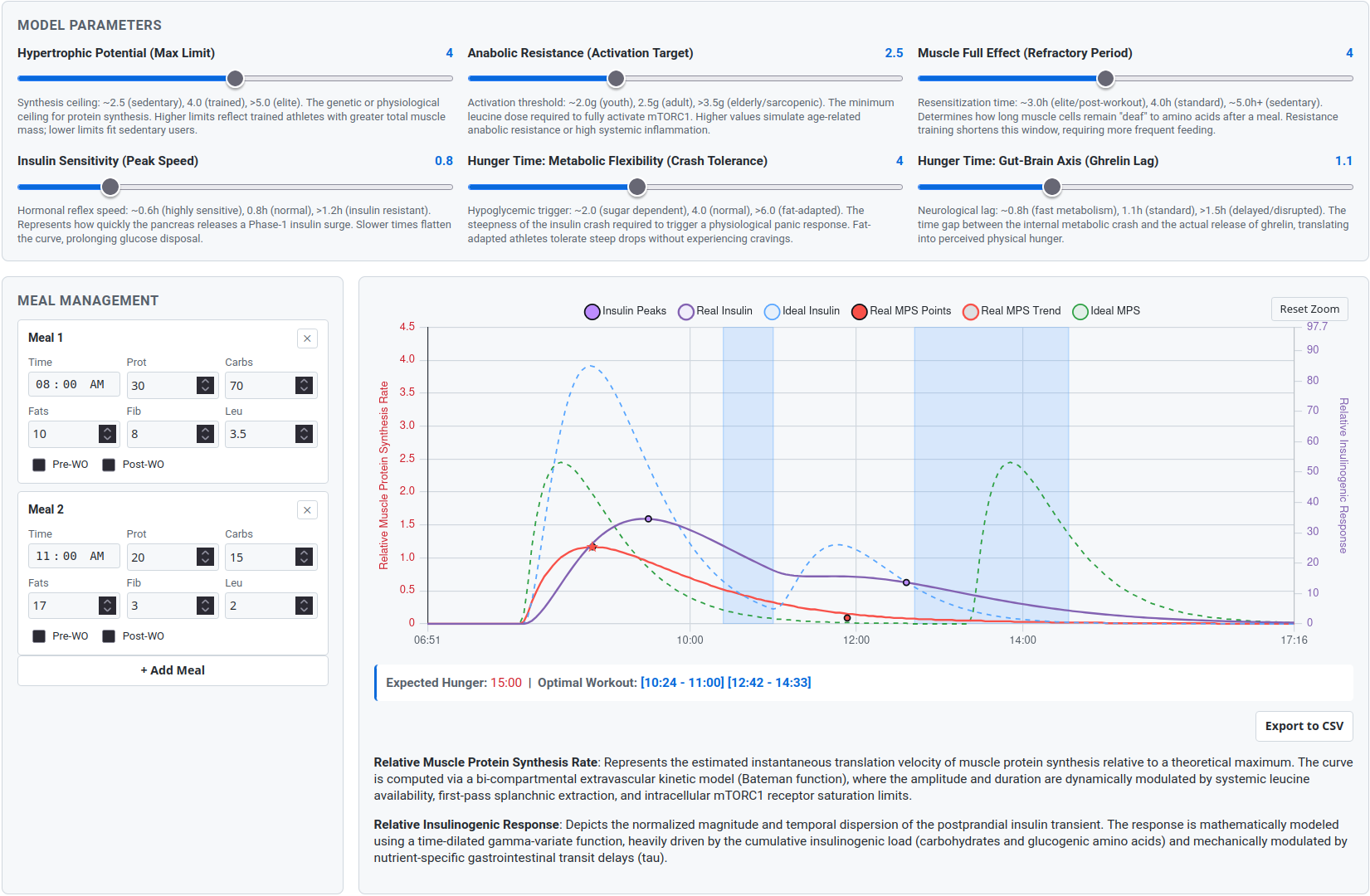}
    \caption{\footnotesize Overview of the MyoBro Dashboard interface, illustrating the physiological parameter sliders and the granular meal management module.}
    \label{fig:dashboard_ui}
\end{figure}

The design of the dashboard is built to compute and visualize physiological outcomes using user-defined constraints and dietary inputs. The interface is divided into three core functional modules:

\begin{itemize}
    \item \textbf{Physiological Parameterization:} The model allows users to adjust key physiological parameters by manipulating sliders, taking into account individual differences. These parameters include:
	\begin{itemize}
\item Hypertrophic Potential: Synthesis ceiling. The genetic or physiological ceiling for protein synthesis. Higher limits are associated with trained athletes with greater total muscle mass, while lower limits are suitable for sedentary individuals. 

\item Anabolic resistance: Activation threshold. The minimum leucine dose required to fully activate mTORC1. Higher values act as a mechanism to simulate age-related anabolic resistance or high systemic inflammation. 

\item Muscle Full Effect (Refractory Period): the period during which the muscle becomes less responsive to stimulation. Measures the duration for which muscle cells remain in a state of "deafness" to amino acids after a meal. Resistance training shortens this window, necessitating a higher frequency of feeding. 

\item Insulin Sensitivity: Hormonal reflex speed. Measures the speed at which the pancreas releases a Phase-1 insulin surge. Slower times flatten the curve, prolonging glucose disposal.

\item Hunger Time: Metabolic Flexibility. Hypoglycemic trigger. The steepness of the insulin-induced crash required to trigger a physiological panic response. Fat-adapted athletes can tolerate steep drops without experiencing cravings. 

\item Hunger Time: Gut-Brain Axis (Ghrelin Lag). Neurological lag. The gap between the internal metabolic crash and the actual release of ghrelin, which translates into perceived physical hunger.

    \end{itemize}
    
    \item \textbf{Nutritional Management:} Meal plans can be created using an easy meal entry process. This entry is characterized by the ability to enter precise meal times, accurate macronutrient breakdown (protein, carbohydrate, fat, and fiber), total leucine intake, and peri-workout status (whether the meal is before or after workout).
    
    \item \textbf{Visual Analytics and Forecasting:} When any of the parameters or inputs is changed, the dashboard generates an updated interactive graph. The graph includes the curve showing the ideal and the real muscle protein synthesis (MPS) rates along with the curves of insulinogenic reactions and insulin spikes. In addition, the underlying mathematical algorithm is able to predict when hunger will come next based on the input meals. Note that the MPS number refers to the actual value of the maximal theoretical saturation of intracellular anabolic signal (mTORC1) resulting from the peak of bioavailable leucine concentration, not the actual mechanical rate of muscle protein synthesis.
\end{itemize}

\begin{figure}[H]
    \centering
    \includegraphics[width=0.95\linewidth]{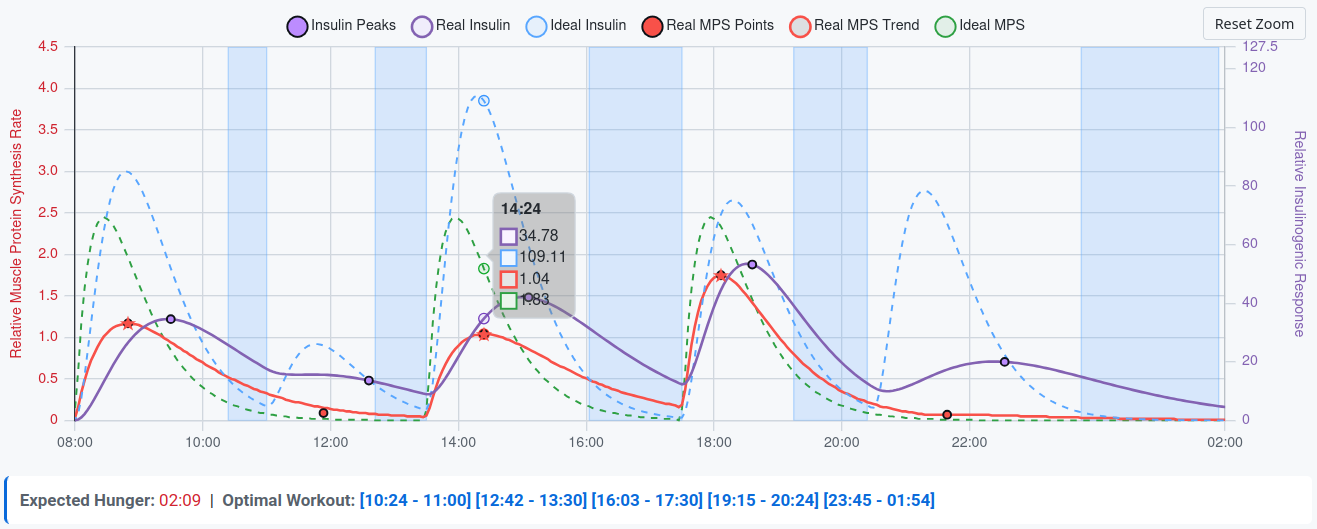}
    \caption{\footnotesize The dashboard features dynamic dual-axis visual output. The chart displays simulated muscle protein synthesis (MPS) rates (left axis) and insulinogenic responses (right axis) over a 24-hour period, which are determined in real time in response to the user's nutritional inputs.}
    \label{fig:dashboard_chart}
\end{figure}

\paragraph{Platform Access and Availability}
To make community testing, empirical evaluation, experiments and review possible, a publicly accessible demonstration instance of the MyoBro Dashboard is available. This public version imposes session limits. Researchers, clinicians, and anyone needing access to intensive data analysis or professional use may request extended sessions directly from the author contacting myobro@myobro.com. 


\section{API: Programmatic Integration}
\label{sec:api_integration}

To facilitate the programmatic integration into external diagnostic tools, third-party applications, and clinical dashboards, the underlying mathematical algorithm is exposed via API. This RESTful endpoint allows developers to calculate complex human physiological states based on a chronological sequence of nutritional inputs and mechanical stimuli.

Access to the API operates on a highly concurrent pay-per-use token architecture. Licensing is strictly granted upon request, catering specifically to researchers, clinical nutritionists, sports scientists, and commercial application developers seeking to embed advanced bio-computational modeling into their software ecosystems. 

\subsection{Endpoint Architecture and Dynamic Configuration}
The core function of the API is to compute curves and predictions over a 24-hour domain. The system evaluates a discrete array of chronological events (meals and workouts) submitted via a JSON payload. 

To account for inter-individual variability (e.g., age, physiological baseline, training history, and general health), the underlying algorithm works on an multidimensional configuration matrix. A subset of parameters is user-modifiable using the the \texttt{config} object in the payload, allowing fine-grained calibration of the methematical model:

\begin{itemize}
    \item \textbf{\texttt{BAT\_MPS\_MAX\_LIMIT}}: Defines the physiological ceiling for muscle protein synthesis (MPS), reflecting hyper-trophic potential (e.g., sedentary vs. elite trained states).
    \item \textbf{\texttt{MPS\_LEUCINE\_THRESHOLD}}: Sets the anabolic resistance activation target. Modulating this variable simulates age-related resistance or systemic inflammation.
    \item \textbf{\texttt{MPS\_IDEAL\_CADENCE}}: The cellular refractory period (Muscle Full Effect), dictating cell resensitization time.
    \item \textbf{\texttt{INS\_T\_BASE\_PEAK}}: Modulates absolute insulin sensitivity and phase-1 reflex speed.
    \item \textbf{\texttt{HUNGER\_THRESHOLD\_DELTA} \& \texttt{HUNGER\_DELAY\_COLLAPSE}}: Configure metabolic flexibility and gut-brain axis lag, dictating tolerance to hypoglycemic crashes and systemic ghrelin release timing.
\end{itemize}

Each macro-nutrient array element contains precise physiological values: proteins (\texttt{p}) and carbohydrates (\texttt{c}) drive MPS potential and insulin surge magnitude respectively; fats (\texttt{f}) and dietary fibers (\texttt{fib}) act as mechanical dampeners, applying dilation to gastric transit curves; and leucine (\texttt{l}) acts as the primary bio-chemical trigger evaluated against the MPS threshold. Boolean flags (\texttt{isPreWorkout}, \texttt{isPostWorkout}) mathematically simulate catecholamine-induced vasoconstriction, which is slowed down by mechanical tension. 

\subsection{Application and Integration Examples}
By extracting predicted outputs, developers can focus on three primary aspects.

\paragraph{Continuous Metabolic Response \& MPS Tracking} The querying applications can reconstitute and display the estimated postprandial pharmacokinetic curve and identify whether the user is able to trigger the anabolic state (denoted by \textit{ok} in the description of the event). The API returns the points of the curves, which are described using a real coordinate array, taking into account the effect of fats and fibers on delayed gastric emptying.
\begin{lstlisting}[language=json, basicstyle=\ttfamily\scriptsize, caption=Metabolic Response Output Snippet]
"mps": {
    "idealData": [{ "x": 7.5, "y": 0.0 }, ...],
    "realCurveData": [{ "x": 7.5, "y": 0.0 }, ...],
    "realPointsData": [{ "x": 7.5, "y": 0, "note": "ok" }],
    "startX": 6,
    "theoretical_next_mps_peak": "11:30"
}
\end{lstlisting}

\paragraph{Workout Window Prediction} Through correlation between anabolic kinetic energy and reduction in insulin transients, the model is able to determine optimal times chronologically for mechanical stimulation, avoiding reactive hypoglycemia during workouts and promoting maximum fat burning and adaptation.
\begin{lstlisting}[language=json, basicstyle=\ttfamily\scriptsize, caption=Workout Window Output Snippet]
"workout": {
  "workout_windows": [
    {
        "start_dec": 9.5,
        "start": "09:30",
        "end_dec": 11.5,
        "end": "11:30"
    }
  ]
}
\end{lstlisting}

\paragraph{Reactive Hypoglycemia and Hunger Forecasting}
Professional dietary applications can identify acute physical hunger in advance. By computing the steepest negative derivative of the blood-glucose clearance curve, the API predicts the moment when systemic ghrelin triggers a neurological crash response. 
\begin{lstlisting}[language=json, basicstyle=\ttfamily\scriptsize, caption=Insulin \& Hunger Output Snippet]
"insulin": {
    "idealCurveData": [...],
    "realCurveData": [...],
    "peaksData": [{ "x": 8.2, "y": 45.6, "mealIndex": 0 }],
    "hungerTime": "10:15"
}
\end{lstlisting}
\subsection{API Performance and Scalability Report}

To assess the scalability and stability of the MyoBro computational API under production-like conditions, I developed a load test that mimics real human behavior. A phased load pattern was implemented, with 50 concurrent virtual users and a delay of 8 to 15 seconds between each request. Such load testing simulates real situations by analyzing metabolic data obtained by a user and then performing another calculation. 

A total of over 1,050 dynamic API requests were executed in five minutes. The system met the reliability standards with a zero failure rate. All validation checks were completed successfully, confirming that all payloads were processed without any issues related to the generation of 500 Internal Server Error responses (database concurrency lock) or malformed JSON responses.

The performance figures for the given computational architecture indicate that response times and scalability exceeded expectations. The mean http\_req\_duration time was exceptionally low at approximately 135 ms. Particularly crucial is the significance of the 95th percentile (P95), which equates to roughly 300 milliseconds, substantially lower than our critical threshold of 3 seconds (3000 milliseconds). Moreover, the maximum http\_req\_duration value was strictly bounded below 750 ms.

\section{Discussion and Conclusions}

This paper introduces a pragmatic algorithm that allows simulating postprandial macronutrient kinetics through mathematical modeling. Using bi-compartmental Bateman functions and FSM logic as alternatives to computationally expensive ODE solvers, the model achieves deterministic execution times in sub-millisecond timescales. Unlike the mechanistic ODE models that have been put forth to describe mTORC1 regulation by leucine \cite{mccoll2024kinetic}, the core of the presented approach relies on a computational tradeoff sacrificing the complex intracellular resolution of multi-pool models in favor of scalability and accessibility for digital health applications. 

In the context of digital health and precision nutrition, there is a clear demand for rapid and scalable physiological assessments. Obtaining continuous empirical data in vivo by using isotopic tracers, venous blood samples, or Continuous Glucose Monitors (CGMs) is invasive, costly, and logistically challenging. The application of a lightweight mathematical engine provides real-time, directional estimates of metabolic states through standard dietary inputs. This approach enables the creation of programmatic tools capable of running instantaneously, which allows obtaining feedback about metabolic states without continuous hardware monitoring and laboratory tests. 

Despite the promising predictive margins (MAPE of 17.2\% to 18.86\%), this algorithm presents some tread-offs. The first limitation is that this system operates as an open-loop prediction algorithm. Human metabolism is a stochastic process, and factors such as psychological stress, hydration, sleep, and microbiome diversity have considerable effects on $\tau$ and insulin sensitivity. Since the algorithm uses only mathematical projections and does not incorporate any real-time corrective data, the system cannot adapt to unpredictable changes in physiology. Secondly, since this model uses macroscopic abstractions, such as reducing the Meal Matrix Effect to a temporal dilation coefficient or simplifying the continuous mTORC1 signaling cascade to a set of discrete FSMs, its predictions are approximation of actual biological processes. Rather than tracking exact biological interactions, the model defines system and time boundaries. 

To conclude, this engine should not be regarded as a replacement for full-fledged clinical diagnostics and medical studies. Instead, it functions as an efficient and accurate computational tool with easy integration capabilities, serving as a neutral and readily applicable bridge between pharmacokinetics and digital health.
\printbibliography

\end{document}